\documentclass[aps, 
 prl, 
 noeprint, 
 superscriptaddress, 
 floatfix, 
 tightenlines, 
 twocolumn
 ]{revtex4-2}

\usepackage[english]{babel}
\usepackage{graphicx} 
\usepackage{listings} 
\usepackage{amsmath,amssymb,amsthm} 
\usepackage[utf8]{inputenc}
\usepackage[toc,page]{appendix}
\usepackage{color}
\usepackage{caption}

\usepackage{hyperref}
\hypersetup{
  colorlinks = true, 
  urlcolor = blue, 
  linkcolor = blue, 
  citecolor = blue 
}

\newcommand{\tx}[1]{\textup{#1}} 
\newcommand{\YO}{Y$_2$O$_3$}

\newcommand{\gtwo}{$g^{(2)} \left ( \tau \right )$}
\newcommand{\gtwoZero}{$g^{(2)} \left ( 0 \right )$}

\begin{document}

\title{Detection of single ions in a nanoparticle coupled to a fiber cavity}
\author{Chetan Deshmukh}
\thanks{These authors contributed equally.}
\author{Eduardo Beattie}
\thanks{These authors contributed equally.}
\author{Bernardo Casabone}
\author{Samuele Grandi}
\email{samuele.grandi@icfo.eu}
\affiliation{ICFO-Institut de Ciencies Fotoniques, The Barcelona Institute of Science and Technology, 08860 Castelldefels (Barcelona), Spain.}
\author{Diana Serrano}
\affiliation{Chimie ParisTech, PSL University, CNRS, Institut de Recherche de Chimie Paris, Paris, France.}
\author{Alban Ferrier}
\affiliation{Chimie ParisTech, PSL University, CNRS, Institut de Recherche de Chimie Paris, Paris, France.}
\affiliation{Faculté des Sciences et Ingénierie, Sorbonne Université, UFR 933, 75005 Paris, France}
\author{Philippe Goldner}
\affiliation{Chimie ParisTech, PSL University, CNRS, Institut de Recherche de Chimie Paris, Paris, France.}
\author{David Hunger}
\affiliation{Karlsruher Institut f\"{u}r Technologie, Physikalisches Institut, Karlsruhe, Germany.}
\affiliation{Karlsruhe Insitute for Technology, Institute for Quantum Materials and Technologies (IQMT), Eggenstein-Leopoldshafen, Germany.}
\author{Hugues de Riedmatten}
\affiliation{ICFO-Institut de Ciencies Fotoniques, The Barcelona Institute of Science and Technology, 08860 Castelldefels (Barcelona), Spain.}
\affiliation{ICREA-Instituci\'o Catalana de Recerca i Estudis Avan\c cats, 08015 Barcelona, Spain.}

\begin{abstract}
Many quantum information protocols require the storage and manipulation of information over long times, and its exchange between nodes of a quantum network across long distances. Implementing these protocols requires an advanced quantum hardware, featuring, for example, a register of long-lived and interacting qubits with an efficient optical interface in the telecommunication band. Here we present the Purcell-enhanced detection of single solid-state ions in erbium-doped nanoparticles placed in a fiber cavity, emitting photons at 1536~nm. The open-access design of the cavity allows for complete tunability both in space and frequency, selecting individual particles and ions. The ions are confined in a volume two orders of magnitude smaller than in previous realizations, increasing the probability of finding ions separated only by a few nanometers which could then interact. We report the detection of individual spectral features presenting saturation of the emission count rate and linewidth, as expected for two-level systems. We also report an uncorrected \gtwoZero\ of 0.24(5) for the emitted field, confirming the presence of a single emitter. Our fully fiber-integrated system is an important step towards the realization of the initially envisioned quantum hardware.
\end{abstract}

\maketitle


\section{Introduction}
The basic constituent element of a distributed quantum computing system has to be able to address and manipulate qubits with long lifetimes and interface them with other computing centers over a quantum network. Such a critical enabling system would likely entail the use of spin-photon interfaces \cite{Awschalom2018}, which allow for the manipulation of long-lived matter qubits and their mapping to photons for long-distance information transfer with low loss. This route has been extensively researched, with systems ranging from the laser trapping of single atoms and ions \cite{Hucul2015,Krutyanskiy2019,VanLeent2022} to solid-state systems like quantum dots \cite{Javadi2018} or colour centres in solids \cite{Bradley2019, Tchebotareva2019}.

To enhance the light-matter interaction and to ensure a proper collection of the photons a common approach is to change the local density of states around the emitter, e.g. by engineering photonic nanostructures or by placing the matter qubit in a low-volume and high-finesse optical cavity. The natural lifetime of the optical transition is reduced through the Purcell effect \cite{Purcell1946}, which could increase the coherence decay rate above external dephasing introduced by interaction with the environment, which for solid-state systems is an especially significant contribution. However, the modifications required for the inclusion of a high-finesse cavity often heavily burden other aspects, for example limiting the maximum number of interconnected qubits or introducing additional decoherence channels due to the nanostructuring of the environment close to the emitters.

\begin{figure*}[t!]
 \centering
 \includegraphics[width=2\columnwidth]{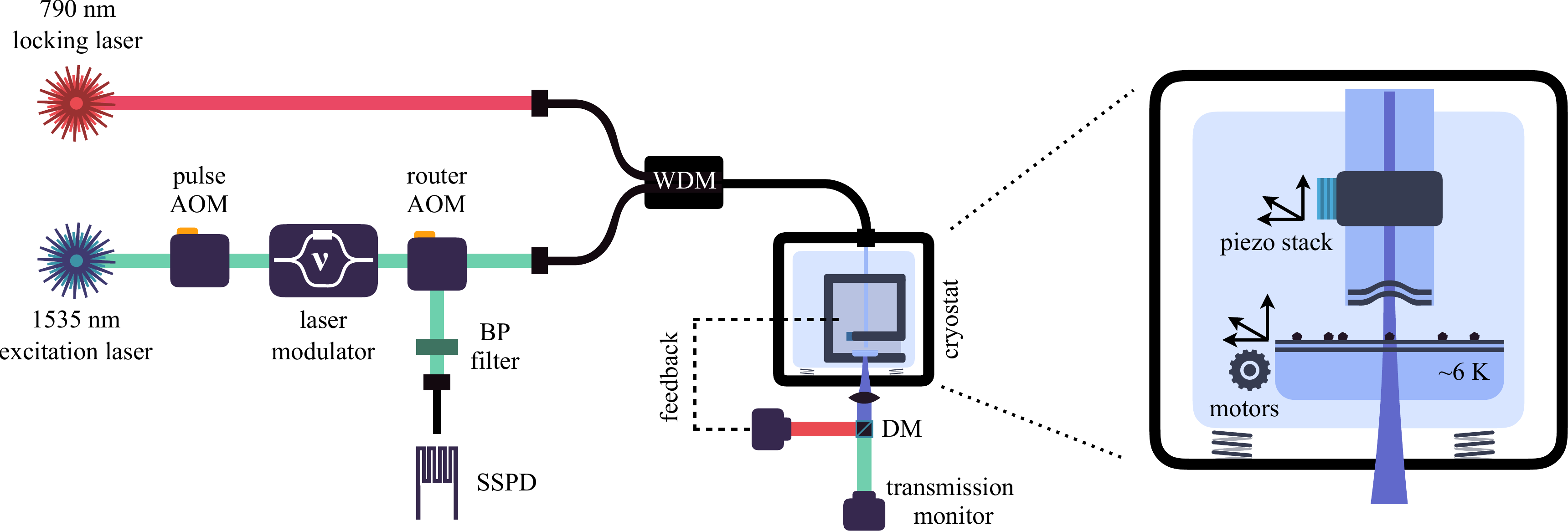}
 \caption{Description of the setup. A laser at 1535~nm provides the excitation light for the erbium ions. The light is shaped into pulses by an acusto-optic modulator (AOM), and its frequency is further controlled using a modulator. The cavity is stabilised using a laser at 790~nm, which is combined with the excitation light through a wavelength-division multiplexer (WDM). The optical fibre is fed into the cryostat where the fibre cavity is placed. The transmission of the 790~nm laser through the cavity is monitored to stabilise the cavity. The fluorescence from the ions is collected back through the optical fibre and directed to superconducting detectors (SSPD) using a router AOM. Right: schematic of the fibre cavity setup. DM: dichroic mirror.}
 \label{fig:setup}
\end{figure*}

Rare earth ion-doped solids (REIDS) in nanostructures provide a system that could benefit from a high Purcell enhancement without many of the associated problems. REIDS have been extensively used as a platform for light-matter interaction, providing an ensemble of atoms with long coherence times naturally trapped in a solid-state matrix \cite{Macfarlane2002}, and they are a leading system for optical quantum memories \cite{deRiedmatten2008,Hedges2010,Rakonjac2021,Ortu2022} and quantum repeater applications \cite{Sangouard2011,Lago-Rivera2021,Liu2021}. Among the rare-earth ions, erbium offers direct operation at telecommunication wavelengths, as well as long optical \cite{Fukumori2020} and spin coherence times \cite{Rancic2018} at low temperature and with suitable magnetic fields.
Recently, efforts have been made to detect single ions, to facilitate the manipulation of individual spins. This approach enables long-lived and efficient spin-photon interfaces and quantum gates between single ion qubits by harnessing dipolar interactions, which open the door to quantum computing \cite{Kinos2021} with a high degree of connectivity both within each node \cite {Kinos2022} and between network nodes. Quantum repeaters using single rare earth-ion qubits have also been proposed, combining a processing ion with long coherence time (e.g. europium \cite{Casabone2018,Serrano2018} or praseodymium \cite{Serrano2019}) and a communication erbium ion at telecommunication wavelength \cite{KimiaeeAsadi2018}. Following early experiments in free space \cite{Kolesov2012,Utikal2014}, single REIs coupled to cavities have emerged as a powerful platform enabling single ion detection and manipulation \cite{Dibos2018,Zhong2018}, single shot read-out \cite{Raha2020,Kindem2020}, addressing of several qubits in the spectral domain \cite{Chen2020,Ulanowski2022}, fast modulation of cavity coupling \cite{Xia2022} and coupling to long-lived nuclear spins \cite{Ruskuc2022}.

Most of these experiments have been done with nanophotonic cavities which have limited tuning capabilities. Experiments coupling erbium ions in a membrane with a high-finesse open-access tunable Fabry-Perot resonator have also been performed, leading to long coherence times and single ion detection \cite{Merkel2020,Ulanowski2022}. Nonetheless, in all the experiments so far the interaction volume was comparable to the cavity mode volume and reached a minimum of order of 0.5~$\lambda^3$. In these conditions, the doping concentration has to remain small to limit the total number of ions such that individual ones can be addressed. This makes challenging to find several ions within a few nm distance, as required for performing dipolar gates \cite{Kinos2022}. 

Here we use an alternative approach exploiting an open-access fiber microcavity and erbium-doped nanoparticles \cite{Casabone2021}. It combines the advantages of an open-access design with the high concentration of dopants available in nanoparticles, all in a fiber-integrated device able to select separate particles and spectrally tune to single solid-state emitters. All the ions are concentrated in a volume more than two orders of magnitude smaller than in previous realizations, due to the low volume of the nanoparticle that is much smaller than the mode of the cavity. We show that Purcell enhancement of up to 120 can be achieved in this setup, and we show clear signatures of single ion detection. Single spectral features are identified and investigated, and we report Zeeman splitting of an emission line in the presence of magnetic field and saturation of the emitted fluorescence for increasing excitation power. Finally, for one of these features we report a measurement of the second-order auto-correlation function \gtwo\ of the collected fluorescence, with an uncorrected zero-delay value well below the limit of 0.5 for single emitters, and compatible with the one expected given the dark count rate of the detectors. The full experiment is run in a fiber-integrated fashion, with both excitation and detection performed through a telecom fiber, and is therefore suitable for long-distance quantum communication.

\section{Description of the Setup}
An outline of the setup is presented in Fig.~\ref{fig:setup}. Our cavity is composed of a planar dielectric mirror, on which the nanoparticles are deposited, and a concave mirror on the tip of a single mode fiber. The latter is realized by first evaporating a spherical indentation on an optical fiber using a CO$_2$ laser, and then depositing a highly-reflective dielectric coating over it \cite{Hunger2010}. The obtained mirror has a radius of curvature of 60~$\mu$m, which, considering a targeted cavity length of 6~$\mu$m, results in a cavity mode waist of 3~$\mu$m and a mode volume $V \approx 40$~$\mu$m$^3$, or about 10~$\lambda^3$. The cavity was designed to maximize the escape of photons towards the optical fibre through the curved fiber mirror, which has a transmittivity of 100~ppm versus the 30~ppm introduced by the planar mirror. The latter coating was additionally engineered to place an antinode of the electric field 50~nm above the surface of the planar mirror, therefore ensuring a maximum field enhancement at the position where nanoparticles are placed. Given these parameters, the empty cavity has an expected finesse of 44,000 at 1535~nm, a quality factor $Q\approx10^5$ and a full-width at half-maximum (FWHM) of 600~MHz.

The cavity was assembled on a home-made nano-positioner placed inside a closed-cycle cryostat, with a design improved from our previous experiment \cite{Casabone2021}. It aims to reduce the mechanical noise of the closed-loop cryostat while at the same time cooling the nanoparticles on the planar mirror. The assembly allows for full 3D positioning of the fiber through the use of piezo-positioners that bend the arm which the fiber is attached to, and for coarse 3D scanning of the position of the planar mirror with three DC motors. This grants total freedom in the selection of the length of the cavity and of the region of the mirror that is analyzed, enabling the free selection of different nanoparticles and optical modes.

The nanoparticles are of \YO\ and are doped with a concentration of 20 ppm of erbium ions. They have a size distribution of 110(30)~nm corresponding to an average volume of $7 \cdot 10^{-4}\, \mu$m$^3$ \cite{Serrano2018}. After synthesis, the nanoparticles are dispersed in ethanol and are deposited on the flat mirror through spin coating \cite{Deshmukh2022}. The mirror is then integrated with the fiber-cavity mount, where it is cooled to about 6.5~K. Fig.~\ref{fig:setup} shows a schematic of the setup. The coating of the mirrors was further designed to provide a double resonance: at 1535~nm, corresponding to the erbium transition, and at 790~nm, which is used to lock the cavity using an additional laser. The cavity is then stabilized to within 20~pm RMS when the cryostat is turned on, mostly limited by the additional mechanical vibrations introduced. This introduced a duty-cycle for our experiment of about $70\%$.

The ions are investigated by addressing the $^4$I$_{15/2}$~$\leftrightarrow$~$^4$I$_{13/2}$ transition at 1535~nm of the erbium ions in the C$_2$-symmetry site and detecting the fluorescence emitted in the cavity. The excitation light is provided by an external-cavity diode laser which is frequency-stabilized to a high-finesse ultra-stable reference cavity. The amplitude of the light is controlled through three double-pass acousto-optic modulators (AOM). In order to scan the frequency over a large range with high frequency resolution and while maintaining the laser locked, we use a frequency modulator developed by Thales Research \& Technology and based on an IQ modulator \cite{Welinski2022}, that allows us to tune the frequency over a range of 3 GHz. With this system we can then control the population of the ions on a timescale much smaller than their lifetime, which allows us to perform a pulsed excitation scheme with excitation and detection of the ions separated in time, removing unwanted fluorescence. Moreover, the signal-to-noise ratio can be increased by reducing the collection window to include only the first portion of the fluorescence exponential decay, at the cost of a reduced total count rate. During the course of the experiment, we used an excitation pulse of 200~$\mu$s and a detection window of 500~$\mu$s, for a repetition rate of 1.4~kHz. The fluorescence emitted by the ions is then collected in the fiber and directed to a superconducting single photon detector, with efficiency $\sim 80\%$ and dark count rate 8~Hz, via a router AOM switch and a narrow-band spectral filter. The efficiency of the superconducting detectors can be reduced to achieve a lower dark count rate, resulting in a higher signal-to-noise ratio.

\begin{figure}[t]
 \centering
 \includegraphics[width=\columnwidth]{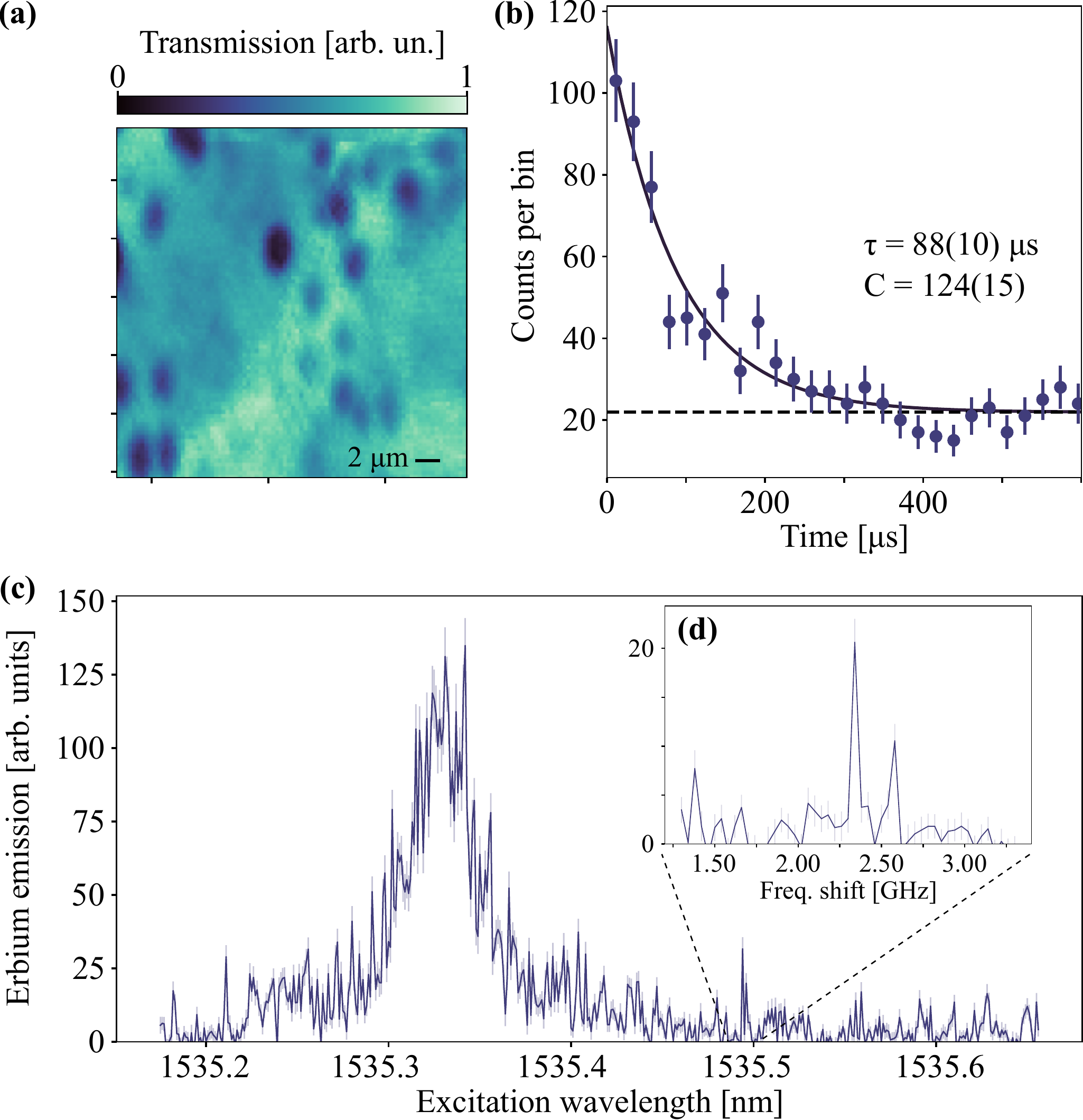}
 \caption{(a) Transmission microscopy image of several nanoparticles, identified by the reduction in cavity transmission. (b) Fluorescence decay measurement of an ensemble of erbium ions in a nanoparticle, showing Purcell enhancement of the lifetime. (c) Fluorescence scan of the inhomogeneous broadening of a nanoparticle, reporting the fluorescence photons collected as the excitation wavelength is changed. (d) Narrow fluorescence scan of the excitation frequency through the laser modulator, where an individual feature can be identified.}
 \label{fig:scan}
\end{figure}

\section{Single Ion Detection}
Individual nanoparticles are located through scattering-loss microscopy using the fiber cavity, detecting the reduction in transmission that is introduced by the particle. A typical scan is shown in Fig.~\ref{fig:scan}(a), showing dark spots where the nanoparticles are located. Suitable particles are chosen based on the loss that they introduce in the cavity, aiming for a minimum reduction in finesse to maintain a high Purcell factor. Fluorescence from the ions is detected through the cavity by first exciting the ions and then opening a detection path to single photon detectors using a router AOM \cite{Casabone2021}. Through fluorescence decay measurements we verified the presence of erbium ions and measured decay constants ranging from one hundred to a few hundred microseconds. Individual values are affected by the orientation of the dipoles in the particle and by the stability of the cavity. Nevertheless, considering the natural lifetime of $\sim 11$~ms \cite{Casabone2021} and a minimum measured lifetime of 88(10)~$\mu$s, reported in Fig.~\ref{fig:scan}(b) for a small ensemble of ions, we were able to demonstrate a Purcell factor $C = 123(14)$. The calculation of the expected Purcell factor requires a value of the branching ratio $\zeta$ of the investigated transition, for which only a lower bound  of 0.13 has been measured so far \cite{Alquedra2022}. Considering that the finesse of the cavity was reduced to about 20'000 while coupled to the particle, then this would correspond to an expected Purcell factor $$C_{exp} = \zeta \frac{3 \lambda ^3}{4 \pi ^2} \frac{Q}{V} > 170$$ which is close to our experimental results. Sub-optimal dipole orientation and cavity instability are among the parameters which could contribute to a reduction of the measured value.

The centre and width of the inhomogeneous broadening of the $^4$I$_{15/2}$ $\leftrightarrow$ $^4$I$_{13/2}$ transition varies among the nanoparticles, and are determined by collecting the erbium fluorescence while performing a wide scan of the excitation light wavelength across several hundreds of pm. One such scan is shown in Fig.~\ref{fig:scan}(c). For this nanoparticle, the width of the inhomogeneous broadening is 6~GHz, and from the reduction in transmission we can estimate its diameter to be about $\sim170$~nm \cite{Casabone2021,Deshmukh2022}. Given its size, an average of 1000~ions in the C2 site ($75\%$ of the total) can be investigated in this nanoparticle in the optical mode of the cavity. We can then deduce a spectral density of $\sim150$~ions/GHz in the nanoparticle, in the centre of the inhomogeneous line \cite{Serrano2019}. The ion density is $400\times 10^3$ $\textup{ions}/\mu$m$^3$. This value is several orders of magnitude higher than in similar systems \cite{Dibos2018, Ulanowski2022}, and of particular importance for the realization of two-qubit gates, for example through ion-ion interaction \cite{Kinos2021,Kinos2022}.

Several individual spectral features can already be recognized in Fig~\ref{fig:scan}(c). Probing their single-emitter nature further, we focused on their spectral linewidth. We observed features with widths ranging from 10~MHz to 400~MHz. Considering the values of homogeneous linewidths in ensembles of nanoparticles measured at the current temperature \cite{Alquedra2022}, we would expect widths in the range of 4 to 10~MHz. However, we also expect that ions closer to the surface of the nanoparticles or close to lattice defects such as crystal boundaries would experience additional dephasing and therefore show wider lines. Some ions showed clear signs of spectral diffusion, ranging from a few seconds to a few minutes \cite{SuppMat}, which could also be compatible with an unfavorable positioning inside the nanoparticle.

\begin{figure}[t]
 \centering
 \includegraphics[width=\columnwidth]{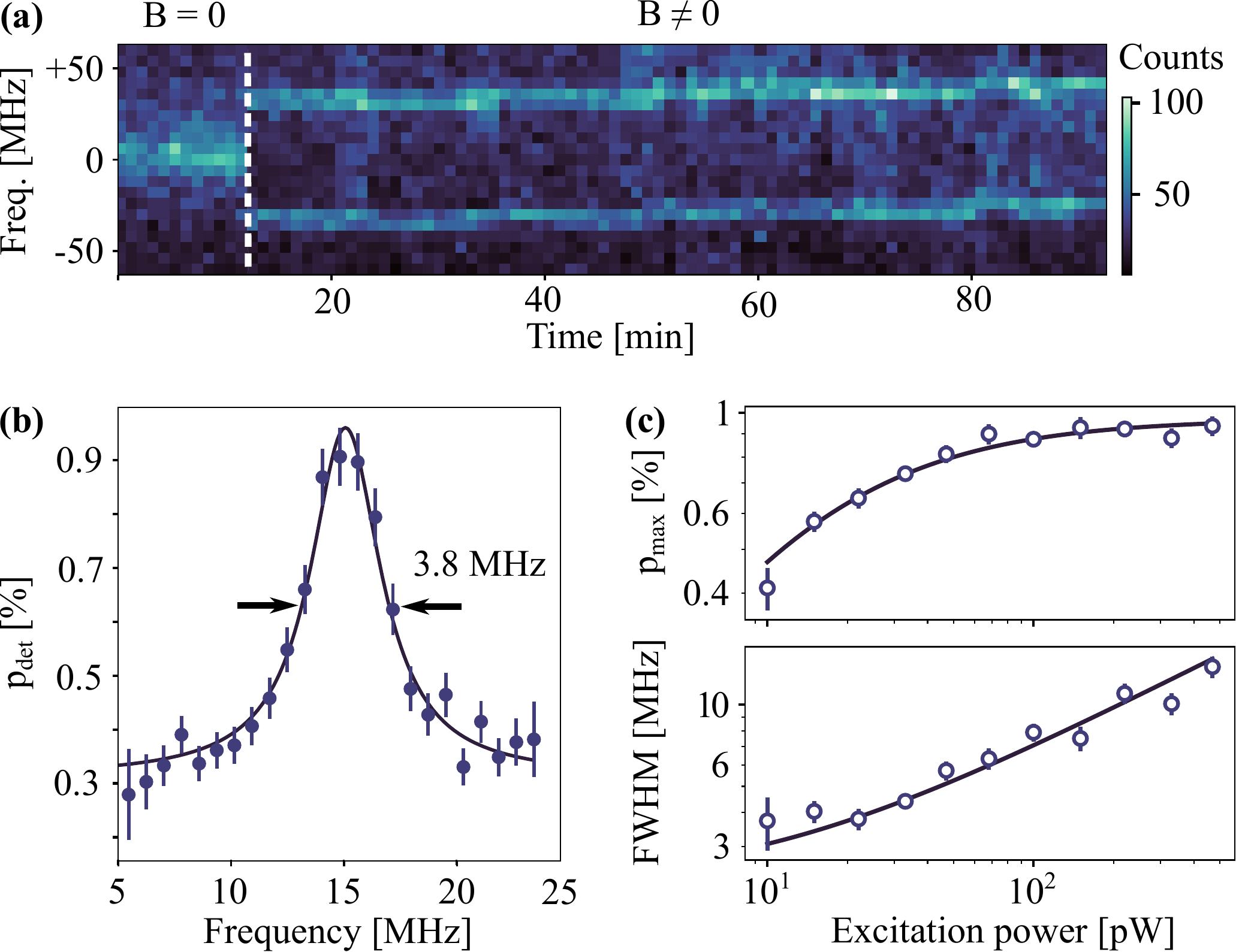}
 \caption{Single ion analysis. (a) Splitting of the fluorescence excitation spectrum after a magnetic field is turned on. A limited drift $<10$~MHz is visible over the course of the measurement. (b) Resonant scan of one of the split lines, obtained for an excitation power of 22~pW in the telecom fiber, showing a FWHM of 3.8~MHz. (c) Saturation scans for linewidth and maximum count rate of a single spectral feature. Both features a non-linear behavior, typical of two-level systems. The solid lines are fits to the expected function, which give a saturation power of 10.7(1)~pW.}
 \label{fig:sats}
\end{figure}

We then turned to the full characterization of a narrow single spectral feature to confirm its single ion nature. We focused on the emitter indicated in Fig.~\ref{fig:scan}(d), for which we measured lifetimes of 350~to~470~$\mu$s, depending on the specific cavity stability conditions \cite{SuppMat}. It showed stable emission, with a variable drift in the central excitation frequency at most comparable to its linewidth \cite{SuppMat}, and no significant long-term drift over more than two months. We investigated further, and by adding a magnetic field with a permanent magnet we were able to fully split the line, as visible in Fig.~\ref{fig:sats}(a), consistent with the Zeeman splitting of a single erbium ion. The spectral linewidth decreases from $\sim30$~MHz to $\sim12$~MHz when applying the magnetic field, which may be due to the presence of a spurious field weakly splitting the levels even when no external field is applied. We then selected one of these lines, and recorded a fluorescence excitation spectrum, reporting a linewidth of 3.8(3)~MHz for low excitation power $P = 22$~pW, as shown in Fig.~\ref{fig:sats}(b). We repeated the measurement for increasing $P$, recording the probability of detection $p_\tx{det}$ of a photon at the centre of the excitation line and its linewidth. We observed saturation of the fluorescence counts compatible with a single ion, as is shown in Fig.~\ref{fig:sats}(c). The first set of data is well fitted by the expression $\frac{S}{S+1}\, p_\tx{max}$, where $S = P/P_\tx{sat}$ is the saturation parameter, $P_\tx{sat}$ is the saturation power and $p_\tx{max}$ is the maximum detection probability. The second set is modeled by the expression $\Delta \nu_0 \sqrt{1+S}$, where $\Delta \nu_0$ is the linewidth at zero power. The data is reported in the bottom panel of Fig.~\ref{fig:sats}(c), and from a fit to these equations we were able to extract $P_\tx{sat}$= 10.7(1)~pW in the telecom fibre, which correspond to $\sim0.05$ intra-cavity photons and $\Delta \nu_0 = 2.20(7)$~MHz. The probability to detect a photon per trial at saturation approaches $p_\tx{max} = 1 \%$, corresponding to a probability that an emitted photon is coupled into the fiber of $p_\tx{f} \approx 4.3 \%$ \cite{SuppMat}.

\begin{figure}[t]
 \centering
 \includegraphics[width=\columnwidth]{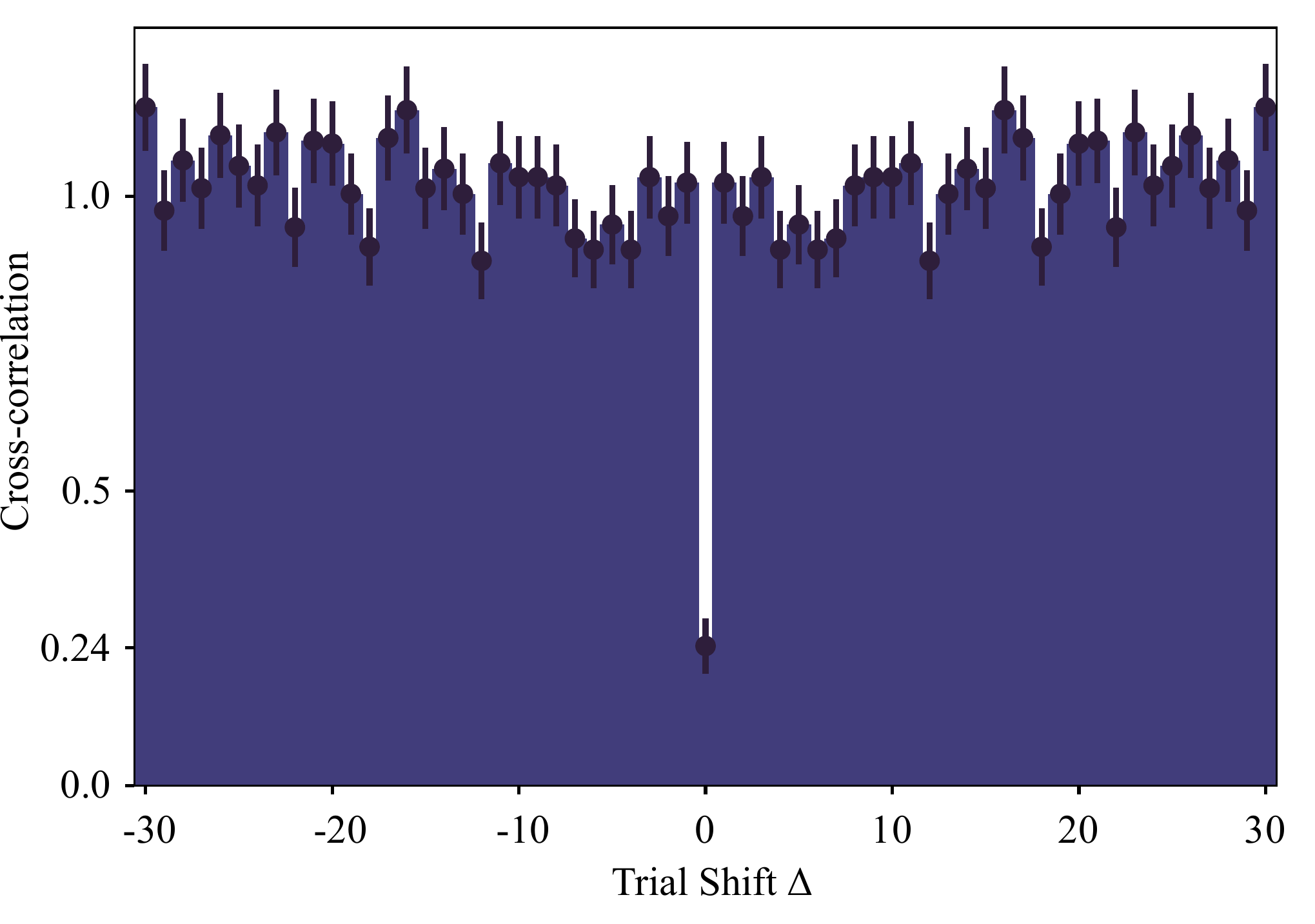}
 \caption{Autocorrelation function of the light collected from the cavity, showing a clear antibunching dip at zero delays, with a value of 0.24(5). The data is not corrected, and the value at zero delays is compatible with the observed signal-to-noise arising from the dark counts of the detectors. These were set to a detection efficiency of 50~$\%$, to reach a dark count rate of 1.4~Hz.}
 \label{fig:g2}
\end{figure}

The non-linear behavior of the detection probability and linewidth for increasing excitation power is a signature of a single two-level system, whereby only one excitation is possible at any given time. Another consequence is the emission of single photons, which can be confirmed by the detection of sub-poissonian statistics in the emitted field. This can be gauged by reconstructing the second-order correlation function $g^{(2)} \left ( \tau \right )$ of the emitted field, also called auto-correlation. The \gtwo\ of the field emitted by a single ion should feature a dip to 0 at $\tau = 0$, presenting at the same time sub-poissonian features and antibunching, both signatures of non-classical light. Any variation from this ideal value comes from a reduced signal-to-noise ratio, due for example to the excitation of neighboring ions or to detector dark counts. We measured the \gtwo\ by using only one detector and measuring the number of photons detected per excitation trial. This was possible due to the low dead-time of our superconducting detectors (50~ns) compared with the temporal length of the photons (350~$\mu$s). By correlating these detection events we were able to reconstruct a \gtwo\, shown in Fig.~\ref{fig:g2}. The data is not background subtracted. The value of $g^{(2)} \left ( 0 \right ) = 0.24(5)$ is well-below the 0.5 threshold for single photon emitters and is compatible with the expected value of a perfect single emitter given the noise counts of our detectors, providing a conclusive proof of the detection of a single erbium ion in a nanoparticle.

\section{Discussion}

Our results represent the first demonstration that a single rare-earth ion can be addressed and detected within a nanoparticle coupled to a fiber-microcavity. In our experiments, hundreds of ions were confined to a volume 2 orders of magnitude smaller than previous realizations, strongly increasing the probability to find ions within a distance that allows strong dipolar interaction. Our platform therefore opens prospects for the realization of quantum processors with hundreds of qubits in a nanoscale volume, which can be efficiently coupled to single photons for quantum networking. Moreover, the erbium ions detected in this work emit single photons at telecommunication wavelength in a fiber-optic setup and could therefore serve as communication qubits. Our approach also enables the coupling of many spectrally-separated ions to the cavity, opening the door to frequency multiplexed quantum nodes \cite{Chen2020, Ulanowski2022}.

There are still several challenges ahead towards these goals. First, the realisation of dipolar ion-ion gates requires long spin coherence times as well as a limited and controllable interaction between neighbouring ion spins. While narrow optical linewidths have been achieved for Er \cite{Fukumori2020,Ourari2023}, spin-spin interactions at high densities would still limit the spin coherence and the fidelity of the quantum gates. Moreover, the anisotropic g-factor of erbium ions prevents the use of dynamical decoupling techniques~\cite{Merkel2021}. However, this interaction could be heavily reduced by employing the nuclear spins of $^{167}$Er combined with strong magnetic fields to freeze out electronic spins, as evidenced by the spin coherence times of 1.3~s that have been measured using these techniques \cite{Rancic2018}. Another possibility would be to use non-Kramers ions with long spin coherence times such as europium~\cite{Casabone2018,Serrano2018} or praseodymium~\cite{Serrano2019}, where the spin-spin interaction is strongly reduced and temperature requirements are less strict. For the latter approach, erbium ions could serve as communication qubits to connect quantum processors, taking advantage of the possibility to co-dope the nanoparticles with several ion species~\cite{Kinos2021}.


A second challenge is to reach the emission of coherent transform-limited photons from the erbium ions to enable the connection of distant nodes via photonic interference. This requires improvements in both the Purcell enhancement and optical coherence time. The first could be significantly increased by improving the finesse and the stability of the cavity. The second is likely limited by the temperature ($>$6K) of the sample, so this could be increased by lowering the operating temperature using a more sophisticated cryostat. Optical homogeneous linewidths of 580~Hz have been achieved recently in Er:Y$_2$O$_3$ ceramics at millikelvin temperature \cite{Fukumori2020}, and while coherence times in nanoparticles are shorter than in ceramics, significant improvement can be expected. An additional benefit of longer coherence times would be the ability to perform coherent excitation and spin initialization, which will readily increase by a factor of four the probability to generate a photon per trial. 

Another parameter that can be improved is the generation and detection efficiency. The probability to couple an emitted photon in the fiber is currently limited by the mode matching between the cavity mode and the fiber, and by intracavity losses due to scattering from the nanoparticles. The former could be improved using fiber cavities with integrated mode matching optics \cite{Gulati2017}, while the latter could be reduced by embedding the nanoparticles, e.g. in a thin layer of \YO.

Open-access microcavities have recently shown great promise. They have been coupled to several solid-state emitters \cite{Riedel2017,Wang2019,Tomm2021}, improving light-matter interaction and photon collection efficiency. Our system adds a fully fiber-coupled interface which could be adapted to other types of nanoparticles or nanocrystals, from other rare-earth ions \cite{Alquedra2022} to colour centres \cite{Benedikter2017} or dye molecules \cite{Pazzagli2018}, or even to crystalline membranes, addressing a register of optically-active solid-state spins.

\section{Acknowledgments}
We thank Sacha Welinski and Perrine Berger from Thales Research \& Technology for the development of the laser modulator. We thank the mechanical workshop at ICFO for help in the fabrication of the cryogenic nanopositioning stage.

This project received funding from the European Union Horizon 2020 research and innovation program within the Flagship on Quantum Technologies through grant 820391 (SQUARE) and under the Marie Sk\l odowska-Curie grant agreement No. 754510 (proBIST), from the Gordon and Betty Moore foundation through Grant GBMF7446 to HdR, from the Government of Spain (PID2019-106850RB-I00 (QRN), Severo Ochoa CEX2019-000910-S, and IJC2020-044956-I (Juan de la Cierva fellowship to SG, funded by MCIN/AEI/10.13039/501100011033)), from MCIN with funding from European Union NextGenerationEU (PRTR-C17.I1), from Fundaci\'o Cellex, Fundaci\'o Mir-Puig, and from Generalitat de Catalunya (CERCA, AGAUR).

%


\end{document}